\input ./style/arxiv-general.cfg
\documentclass[aoas,MSNbibl,nameyear,seceqn,dvips]{arximspdf}
\makeatletter
   \@ifpackageloaded{graphicx}{}{\usepackage{graphicx}}
\makeatother
\usepackage{framed}
\usepackage{dcolumn}
\usepackage{url,breakurl}

\doi{10.1214/15-AOAS848}
\volume{9}
\issue{3}
\pubyear{2015}
\firstpage{1350}
\lastpage{1371}
\docsubty{FLA}

\makeatletter
\newcolumntype{d}[1]{D{.}{.}{#1}}
\newcommand{\eqref}[1]{(\ref{#1})}
\makeatother

\begin{document}
\begin{frontmatter}

\title{Interpretable classifiers using rules and Bayesian analysis:
Building a better stroke prediction model}
\runtitle{Interpretable classifiers using rules and Bayesian analysis}

\begin{aug}
\author[A]{\fnms{Benjamin} \snm{Letham}\corref{}\thanksref{m1,T1}\ead[label=e1]{bletham@mit.edu}},
\author[B]{\fnms{Cynthia} \snm{Rudin}\thanksref{m1,T1}\ead[label=e2]{rudin@mit.edu}},
\author[C]{\fnms{Tyler H.} \snm{McCormick}\thanksref{m2,T2}\ead[label=e3]{tylermc@u.washington.edu}}\\
\and
\author[D]{\fnms{David} \snm{Madigan}\thanksref{m3,T3}\ead[label=e4]{madigan@stat.columbia.edu}}
\runauthor{Letham, Rudin, McCormick and Madigan}
\thankstext{T1}{Supported in part by NSF
CAREER Grant IIS-1053407 from the
National Science Foundation to C.~Rudin, and awards from Siemens and
Wistron.}
\thankstext{T2}{Supported in part by a Google Faculty Award
and NIAID Grant R01 HD54511.}
\thankstext{T3}{Supported in part by Grant
R01 GM87600-01 from the National Institutes of Health.}
\affiliation{Massachusetts Institute of Technology\thanksmark{m1},
University of Washington\thanksmark{m2}\\ and Columbia
University\thanksmark{m3}}
\address[A]{B. Letham\\
Operations Research Center\\
Massachusetts Institute of Technology\\
Cambridge, Massachusetts 02139\\
USA\\
\printead{e1}}
\address[B]{C. Rudin\\
Computer Science and\\
\quad Artificial Intelligence Laboratory\\
Massachusetts Institute of Technology\\
Cambridge, Massachusetts 02139\\
USA\\
\printead{e2}}
\address[C]{T. H. McCormick\\
Department of Statistics\\
Department of Sociology\\
University of Washington\\
Seattle, Washington 98105\\
USA\\
\printead{e3}}
\address[D]{D. Madigan\\
Department of Statistics\\
Columbia University\\
New York, New York 10027\\
USA\\
\printead{e4}\hspace*{38pt}}
\end{aug}

%
\received{\smonth{10} \syear{2013}}
%
\revised{\smonth{4} \syear{2015}}

%
\begin{abstract}
We aim to produce predictive models that are not only accurate, but are
also interpretable to human experts. Our models are decision lists,
which consist of a series of \emph{if}\ldots\emph{then}\ldots statements
(e.g., \emph{if high blood pressure, then stroke}) that discretize a
high-dimensional, multivariate feature space into a series of simple,
readily interpretable decision statements. We introduce a generative
model called Bayesian Rule Lists that yields a posterior distribution
over possible decision lists. It employs a novel prior structure to
encourage sparsity. Our experiments show that Bayesian Rule Lists has
predictive accuracy on par with the current top algorithms for
prediction in machine learning. Our method is motivated by recent
developments in personalized medicine, and can be used to produce
highly accurate and interpretable medical scoring systems. We
demonstrate this by producing an alternative to the CHADS$_2$ score,
actively used in clinical practice for estimating the risk of stroke in
patients that have atrial fibrillation. Our model is as interpretable
as CHADS$_2$, but more accurate.
\end{abstract}

%
\end{frontmatter}

\section{Introduction}\label{sec:intro}

Our goal is to build predictive models that are highly accurate, yet
are highly interpretable. These predictive models will be in the form
of sparse \textit{decision lists}, which consist of a series of \emph
{if\ldots then\ldots} statements where the \emph{if} statements define a
partition of a set of features and the \emph{then} statements
correspond to the predicted outcome of interest.
Because of this form, a decision list model naturally provides a reason
for each prediction that it makes.
Figure~\ref{fig:titanic} presents an example decision list that we
created using the Titanic data set available in R. This data set
provides details about each passenger on the Titanic, including whether
the passenger was an adult or child, male or female, and their class
(1st, 2nd, 3rd or crew). The goal is to predict whether the passenger
survived based on his or her features. The list provides an explanation
for each prediction that is made. For example, we predict that a
passenger is less likely to survive than not \emph{because} he or she
was in the 3rd class. The list in Figure~\ref{fig:titanic} is one
accurate and interpretable decision list for predicting survival on the
Titanic, possibly one of many such lists. Our goal is to learn these
lists from data.

\begin{figure}
\begin{framed}
\begin{flushleft}
\textbf{if} male \textbf{and} adult \textbf{then} \textit
{survival probability} 21\% (19\%--23\%)\\
\textbf{else if} 3rd class \textbf{then} \textit{survival probability}
44\% (38\%--51\%)\\
\textbf{else if} 1st class \textbf{then} \textit{survival probability}
96\% (92\%--99\%)\\
\textbf{else} \textit{survival probability} 88\% (82\%--94\%)
\end{flushleft}
\end{framed}
\caption{Decision list for Titanic. In parentheses is the 95\% credible
interval for the survival probability.}
\label{fig:titanic}
\end{figure}

Our model, called Bayesian Rule Lists (BRL), produces a posterior
distribution over permutations of \emph{if\ldots then\ldots} rules, starting
from a large, pre-mined set of possible rules. The decision lists with
high posterior probability tend to be both accurate and interpretable,
where the interpretability comes from a hierarchical prior over
permutations of rules. The prior favors concise decision lists that
have a small number of total rules, where the rules have few terms in
the left-hand side.

BRL provides a new type of balance between accuracy, interpretability
and computation. Consider the challenge of constructing a predictive
model that discretizes the input space in the same way as decision
trees [\citet{Breiman84,Quinlan93}], decision lists [\citet{Rivest87}] or
associative classifiers [\citet{Liu98}]. Greedy construction methods
like classification and regression trees (CART) or C5.0 are not
particularly computationally demanding, but, in practice, the
greediness heavily affects the quality of the solution, both in terms
of accuracy and interpretability. At the same time, optimizing a
decision tree over the full space of all possible splits is not a
tractable problem. BRL strikes a balance between these extremes, in
that its solutions are not constructed in a greedy way involving
splitting and pruning, yet it can solve problems at the scale required
to have an impact in real problems in science or society, including
modern healthcare.

A major source of BRL's practical feasibility is the fact that it uses
pre-mined rules, which reduces the model space to that of permutations
of rules as opposed to all possible sets of splits. The complexity of
the problem then depends on the number of pre-mined rules rather than
on the full space of feature combinations; in a sense, this algorithm
scales with the sparsity of the data set rather than the number of
features. As long as the pre-mined set of rules is sufficiently
expressive, an accurate decision list can be found and, in fact, the
smaller model space might improve generalization [through the
lens of statistical learning theory, \citet{Vapnik95}]. An additional
advantage to using pre-mined rules is that each rule is independently
both interpretable and informative about the data.

BRL's prior structure encourages decision lists that are sparse. Sparse
decision lists serve the purpose of not only producing a more
interpretable model, but also reducing computation, as most of the
sampling iterations take place within a small set of permutations
corresponding to the sparse decision lists. In practice, BRL is able to
compute predictive models with accuracy comparable to state-of-the-art
machine learning methods, yet maintain the same level of
interpretability as medical scoring systems.

The motivation for our work lies in developing interpretable
patient-level predictive models using massive observational medical
data. To this end, we use BRL to construct an alternative to the
CHADS$_2$ score of \citet{Gage:2001ed}. CHADS$_2$ is widely used in
medical practice to predict stroke in patients with atrial
fibrillation. A patient's CHADS$_2$ score is computed by assigning one
``point'' each for the presence of congestive heart failure (C),
hypertension (H), age 75 years or older (A) and diabetes mellitus (D),
and by assigning 2 points for history of stroke, transient ischemic
attack or thromoembolism (S$_2$). The CHADS$_{2}$ score considers only
5 factors, whereas the updated CHA$_{2}$DS$_{2}$-VASc score [\citet
{Lib10}] includes three additional risk factors: vascular disease (V),
age 65 to 74 years old (A) and female gender (Sc). Higher scores
correspond to increased risk. In the study defining the CHADS$_2$ score
[\citet{Gage:2001ed}], the score was calibrated with stroke risks using
a database of 1733 Medicare beneficiaries followed for, on average,
about a year.

Our alternative to the CHADS$_2$ was constructed using 12,586 patients
and 4148 factors. Because we are using statistical learning, we are
able to consider significantly more features; this constitutes over
6000 times the amount of data used for the original CHADS$_2$ study. In
our experiments we compared the stroke prediction performance of BRL to
CHADS$_2$ and CHA$_{2}$DS$_{2}$-VASc, as well as to a collection of
state-of-the-art machine learning algorithms: C5.0 [\citet{Quinlan93}],
CART [\citet{Breiman84}], $\ell_1$-regularized logistic regression,
support vector machines [\citet{Vapnik95}], random forests [\citet
{Breiman01}], and Bayesian CART [\citet{Dension:1998hl,Chipman:1998jh}].
The balance of accuracy and interpretability obtained by BRL is not
easy to obtain through other means: None of the machine learning
methods we tried could obtain both the same level of accuracy and the
same level of interpretability.

\section{Bayesian rule lists}\label{sec:method}
The setting for BRL is multi-class classification, where the set of
possible labels is $1,\ldots,L$. In the case of predicting stroke risk,
there are two labels: stroke or no stroke.
The training data are pairs $\{(x_i,y_i)\}_{i=1}^n$, where $x_i\in
\mathbb{R}^d$ are the features of observation $i$, and $y_i$ are the
labels, $y_i \in\{1,\ldots,L\}$. We let $\mathbf{x} = (x_1,\ldots
,x_n)$ and $\mathbf{y} = (y_1,\ldots,y_n)$.

In Sections~\ref{sec:bayesiandls} and \ref{sec:mining} we provide the
association rule concepts and notation upon which the method is built.
Section~\ref{sec:generative} introduces BRL by outlining the generative
model. Sections~\ref{sec:prior} and \ref{sec:likelihood} provide
detailed descriptions of the prior and likelihood, and then Sections~\ref{sec:mcmc}
and \ref{sec:posterior} describe sampling and posterior
predictive distributions.

\subsection{Bayesian association rules and Bayesian decision
lists}\label{sec:bayesiandls}
An association rule $a\rightarrow b$ is an implication with an
antecedent $a$ and a consequent $b$. For the purposes of
classification, the antecedent is an assertion about the feature vector
$x_i$ that is either true or false, for example, ``$x_{i,1} = 1 \mbox
{ and } x_{i,2} = 0$.'' This antecedent contains two conditions, which
we call the cardinality of the antecedent. The consequent $b$ would
typically be a predicted label $y$. A Bayesian association rule has a
multinomial distribution over labels as its consequent rather than a
single label:
\[
a \rightarrow y \sim\operatorname{Multinomial}(\bolds{\theta)}.
\]
The multinomial probability is then given a prior, leading to a \textit
{prior consequent distribution}:
\[
\bolds{\theta} | \bolds{\alpha} \sim\operatorname{Dirichlet}(\bolds{\alpha}).
\]
Given observations $(\mathbf{x},\mathbf{y})$ classified by this rule,
we let $N_{\cdot,l}$ be the number of observations with label $y_i =
l$, and $N = (N_{\cdot,1},\ldots,N_{\cdot,L})$. We then obtain a \textit
{posterior consequent distribution}:
\[
\bolds{\theta} | \mathbf{x},\mathbf{y},\bolds{\alpha} \sim\operatorname {Dirichlet}(
\bolds{\alpha}+N).
\]

The core of a Bayesian decision list is an ordered antecedent list $d =
(a_1,\ldots,a_m)$. Let $N_{j,l}$ be the number of observations $x_i$
that satisfy $a_j$ but not any of $a_1,\ldots,a_{j-1}$, and that have
label $y_i=l$. This is the number of observations to be classified by
antecedent $a_j$ that have label $l$. Let $N_{0,l}$ be the number of
observations that do not satisfy any of $a_1,\ldots,a_m$ and that have
label $l$. Let $\mathbf{N}_j = (N_{j,1},\ldots,N_{j,L})$ and $\mathbf
{N} = (\mathbf{N}_0,\ldots,\mathbf{N}_m)$.

A Bayesian decision list $D = (d,\bolds{\alpha},\mathbf{N})$ is an
ordered list of antecedents together with their posterior consequent
distributions. The posterior consequent distributions are obtained by
excluding data that have satisfied an earlier antecedent in the list. A
Bayesian decision list then takes the form:\vspace*{6pt}
\begin{flushleft}
\textbf{if} $a_1$ \textbf{then} $y \sim\operatorname
{Multinomial}(\bolds{\theta}_1)$, $\bolds{\theta}_1 \sim\operatorname
{Dirichlet}(\bolds{\alpha}+\mathbf{N}_1)$\\
\textbf{else if} $a_2$ \textbf{then} $y \sim\operatorname
{Multinomial}(\bolds{\theta}_2)$, $\bolds{\theta}_2 \sim\operatorname
{Dirichlet}(\bolds{\alpha}+\mathbf{N}_2)$\\
$\vdots$\\
\textbf{else if} $a_m$ \textbf{then} $y \sim\operatorname
{Multinomial}(\bolds{\theta}_m)$, $\bolds{\theta}_m \sim\operatorname
{Dirichlet}(\bolds{\alpha}+\mathbf{N}_m)$\\
\textbf{else} $y \sim\operatorname{Multinomial}(\bolds{\theta}_{0})$,
$\bolds{\theta}_{0} \sim\operatorname{Dirichlet}(\bolds{\alpha}+\mathbf{N}_0)$.\vspace*{6pt}
\end{flushleft}
Any observations that do not satisfy any of the antecedents in $d$ are
classified using the parameter $\theta_0$, which we call the default
rule parameter.

\subsection{Antecedent mining}\label{sec:mining}
We are interested in forming Bayesian decision lists whose antecedents
are a subset of a preselected collection of antecedents. For data with
binary or categorical features this can be done using frequent itemset
mining, where itemsets are used as antecedents. In our experiments, the
features were binary and we used the FP-Growth algorithm [\citet
{Borgelt05}] for antecedent mining, which finds all itemsets that
satisfy constraints on minimum support and maximum cardinality. This
means each antecedent applies to a sufficiently large amount of data
and does not have too many conditions. For binary or categorical
features the particular choice of the itemset mining algorithm is
unimportant, as the output is an exhaustive list of all itemsets
satisfying the constraints. Other algorithms, such as Apriori or Eclat
[\citet{Agrawal94,Zaki00}], would return an identical set of
antecedents as FP-Growth if given the same minimum support and maximum
cardinality constraints. Because the goal is to obtain decision lists
with few rules and few conditions per rule, we need not include any
itemsets that apply only to a small number of observations or have a
large number of conditions. Thus, frequent itemset mining allows us to
significantly reduce the size of the feature space, compared to
considering all possible combinations of features.

The frequent itemset mining that we do in our experiments produces only
antecedents with sets of features, such as ``diabetes and heart
disease.'' Other techniques could be used for mining antecedents with
negation, such as ``not diabetes'' [\citet{Wu04}]. For data with
continuous features, a variety of procedures exist for antecedent
mining [\citet{Fayyad93,Dougherty95,Srikant96}]. Alternatively, one
can create categorical features using interpretable thresholds (e.g.,
ages 40--49, 50--59, etc.) or interpretable quantiles (e.g., quartiles)---we took this approach in our experiments.

We let $\mathcal{A}$ represent the complete, pre-mined collection of
antecedents, and suppose that $\mathcal{A}$ contains $|\mathcal{A}|$
antecedents with up to $C$ conditions in each antecedent.

\subsection{Generative model}\label{sec:generative}
We now sketch the generative model for the labels $\mathbf{y}$ from the
observations $\mathbf{x}$ and antecedents $\mathcal{A}$. Define
$a_{<j}$ as the antecedents before $j$ in the rule list if there are
any, \textit{for example}, $a_{<3} = \{a_1, a_2\}$. Similarly, let
$c_j$ be the cardinality of antecedent $a_j$, and $c_{<j}$ the
cardinalities of the antecedents before $j$ in the rule list. The
generative model is then:
\begin{itemize}
\item[--] Sample a decision list length $m \sim p(m|\lambda)$.
\item[--] Sample the default rule parameter $\theta_0 \sim\operatorname
{Dirichlet}(\alpha)$.
\item[--] For decision list rule $j=1,\ldots,m$:

\item[\quad ] Sample the cardinality of antecedent $a_j$ in $d$ as $c_j \sim
p(c_j|c_{<j}, \mathcal{A}, \eta)$.

\item[\quad ] Sample $a_j$ of cardinality $c_j$ from
$p(a_j|a_{<j},c_j,\mathcal{A})$.

\item[\quad ] Sample rule consequent parameter $\theta_j \sim\operatorname
{Dirichlet}(\alpha)$.

\item[--] For observation $i=1,\ldots,n$:

\item[\quad ] Find the antecedent $a_j$ in $d$ that is the first that
applies to $x_i$.
\item[\quad ] If no antecedents in $d$ apply, set $j=0$.
\item[\quad ] Sample $y_i \sim\operatorname{Multinomial}(\theta_j)$.
\end{itemize}

Our goal is to sample from the posterior distribution over antecedent lists:
\[
p(d|\mathbf{x},\mathbf{y},\mathcal{A},\bolds{\alpha},\lambda,\eta) \propto p(
\mathbf{y}|\mathbf{x},d,\bolds{\alpha}) p(d|\mathcal {A},\lambda,\eta).
\]
Given $d$, we can compute the posterior consequent distributions
required to construct a Bayesian decision list as in Section~\ref{sec:bayesiandls}. Three prior hyperparameters must be specified by the
user: $\bolds{\alpha}$, $\lambda$ and $\eta$. We will see in Sections~\ref{sec:prior} and \ref{sec:likelihood} that these hyperparameters
have natural interpretations that suggest the values to which they
should be set.

\subsection{The hierarchical prior for antecedent lists}\label{sec:prior}
Suppose the list of antecedents $d$ has length $m$ and antecedent
cardinalities $c_1,\ldots,c_m$. The prior probability of $d$ is defined
hierarchically as
%
\begin{equation}
\label{eq:priorprob} p(d|\mathcal{A},\lambda,\eta) = p(m|\mathcal{A},\lambda) \prod
_{j=1}^m p(c_j|c_{<j},
\mathcal{A},\eta) p(a_j|a_{<j},c_j,\mathcal{A}).
\end{equation}
We take the distributions for list length $m$ and antecedent
cardinality $c_j$ to be Poisson with parameters $\lambda$ and $\eta$,
respectively, with proper truncation to account for the finite number
of antecedents in $\mathcal{A}$. Specifically, the distribution of $m$
is Poisson truncated at the total number of preselected antecedents:
\[
p(m|\mathcal{A},\lambda) = \frac{(\lambda^m/m!)}{\sum_{j=0}^{|\mathcal
{A}|} (\lambda^j/j!)},\qquad m=0,\ldots,|\mathcal{A}|.
\]
This truncated Poisson is a proper prior, and is a natural choice
because of its simple parameterization. Specifically, this prior has
the desirable property that when $|\mathcal{A}|$ is large compared to
the desired size of the decision list, as will generally be the case
when seeking an interpretable decision list, the prior expected
decision list length $\mathbb{E}[m|\mathcal{A},\lambda]$ is
approximately equal to $\lambda$. The prior hyperparameter $\lambda$
can then be set to the prior belief of the list length required to
model the data. A Poisson distribution is used in a similar way in the
hierarchical prior of \citet{Wu:2007dh}.

The distribution of $c_j$ must be truncated at zero and at the maximum
antecedent cardinality $C$. Additionally, any cardinalities that have
been exhausted by point $j$ in the decision list sampling must be
excluded. Let $R_j(c_1,\ldots,c_j,\mathcal{A})$ be the set of
antecedent cardinalities that are available after drawing antecedent
$j$. For example, if $\mathcal{A}$ contains antecedents of size $1$,
$2$ and $4$, then we begin with $R_0(\mathcal{A}) = \{1,2,4\}$. If
$\mathcal{A}$ contains only $2$ rules of size $4$ and $c_1=c_2=4$, then
$R_2(c_1,c_2,\mathcal{A}) = \{1,2\}$ as antecedents of size $4$ have
been exhausted. We now take $p(c_j|c_{<j},\mathcal{A},\eta) $ as
Poisson truncated to remove values for which no rules are available
with that cardinality:
\[
p(c_j|c_{<j},\mathcal{A},\eta) = \frac{(\eta^{c_j}/c_j!)}{\sum_{k \in
R_{j-1}(c_{<j},\mathcal{A})} (\eta^k/k!)},\qquad
c_j \in R_{j-1}(c_{<j},\mathcal{A}).
\]
If the number of rules of different sizes is large compared to $\lambda
$, and $\eta$ is small compared to $C$, the prior expected average
antecedent cardinality is close to $\eta$. Thus, $\eta$ can be set to
the prior belief of the antecedent cardinality required to model the data.

Once the antecedent cardinality $c_j$ has been selected, the antecedent
$a_j$ must be sampled from all available antecedents in $\mathcal{A}$
of size $c_j$. Here, we use a uniform distribution over antecedents in
$\mathcal{A}$ of size $c_j$, excluding those in $a_{<j}$:
%
\begin{equation}
\label{eq:uniformprior} p(a_j|a_{<j},c_j,\mathcal{A})
\propto1,\qquad a_j \in\bigl\{a \in\mathcal {A} \setminus a_{<j}: |a| =
c_j\bigr\}.
\end{equation}
It is straightforward to sample an ordered antecedent list $d$ from the
prior by following the generative model, using the provided distributions.

\subsection{The likelihood function}\label{sec:likelihood}
The likelihood function follows directly from the generative model. Let
$\bolds{\theta} = (\theta_0,\theta_1,\ldots,\theta_m)$ be the
consequent parameters corresponding to each antecedent in $d$, together
with the default rule parameter $\theta_0$. Then, the likelihood is the
product of the multinomial probability mass functions for the observed
label counts at each rule:
\[
p(\mathbf{y}| \mathbf{x},d,\bolds{\theta})= \prod_{j: \sum_l N_{j,l}
> 0}
\operatorname{Multinomial}(\mathbf{N}_j | \theta_j),
\]
with
\[
\theta_j \sim\operatorname{Dirichlet}(\bolds{\alpha}).
\]
We can marginalize over $\theta_j$ in each multinomial distribution in
the above product, obtaining, through the standard derivation of the
Dirichlet-\break multinomial distribution,
\begin{eqnarray*}
p(\mathbf{y}|\mathbf{x},d,\bolds{\alpha})&=&\prod_{j=0}^m
\frac{\Gamma(
\sum_{l=1}^{L}\alpha_l)}{ \Gamma( \sum_{l=1}^{L} N_{j,l} + \alpha
_l)}\times\prod_{l=1}^{L}
\frac{ \Gamma(N_{j,l}+\alpha_l)}{ \Gamma
(\alpha_l)}
\\
&&{}\propto\prod_{j=0}^m \frac{ \prod_{l=1}^{L} \Gamma(N_{j,l}+\alpha
_l)}{ \Gamma( \sum_{l=1}^{L} N_{j,l} + \alpha_l)}.
\end{eqnarray*}

The prior hyperparameter $\bolds{\alpha}$ has the usual Bayesian
interpretation of pseudocounts. In our experiments, we set $\alpha_l =
1$ for all $l$, producing a uniform prior. Other approaches for setting
prior hyperparameters such as empirical Bayes are also applicable.

\subsection{Markov chain Monte Carlo sampling}\label{sec:mcmc}
We do Metropolis--Hastings sampling of $d$, generating the proposed
$d^*$ from the current $d^t$ using one of three options: (1) Move an
antecedent in $d^t$ to a different position in the list. (2) Add an
antecedent from $\mathcal{A}$ that is not currently in $d^t$ into the
list. (3) Remove an antecedent from $d^t$. Which antecedents to adjust
and their new positions are chosen uniformly at random at each step.
The option to move, add or remove is also chosen uniformly. The
probabilities for the proposal distribution $Q(d^*|d^t)$ depend on the
size of the antecedent list, the number of pre-mined antecedents, and
whether the proposal is a move, addition or removal. For the uniform
distribution that we used, the proposal probabilities for a $d^*$
produced by one of the three proposal types is
\[
Q\bigl(d^*|d^t,\mathcal{A}\bigr) = %
\cases{
\displaystyle\frac{1}{(|d^t|)(|d^t|-1)}, &\quad $\mbox{if move proposal}$,\vspace*{2pt}
\cr
\displaystyle\frac{1}{(|\mathcal{A}| - |d^t|)(|d^t|+1)},
&\quad $\mbox{if add proposal}$,\vspace*{2pt}
\cr
\displaystyle\frac{1}{|d^t|}, &\quad $\mbox{if
remove proposal}$.} %
\]
To explain these probabilities, if there is a move proposal, we
consider the number of possible antecedents to move and the number of
possible positions for it; if there is an add proposal, we consider the
number of possible antecedents to add to the list and the number of
positions to place a new antecedent; for remove proposals we consider
the number of possible antecedents to remove. This sampling algorithm
is related to those used for Bayesian Decision Tree models
[\citeauthor{Chipman:2002hc} (\citeyear{Chipman:1998jh,Chipman:2002hc}), \citet{Wu:2007dh}] and to methods for
exploring tree spaces [\citet{Madigan11}].

For every MCMC run, we ran 3 chains, each initialized independently
from a random sample from the prior. We discarded the first half of
simulations as burn-in, and then assessed chain convergence using the
Gelman--Rubin convergence diagnostic applied to the log posterior
density [\citet{Gelman92}]. We considered chains to have converged when
the diagnostic $\hat{R}<1.05$.

\subsection{The posterior predictive distribution and point
estimates}\label{sec:posterior}
Given the posterior $p(d|\mathbf{x},\mathbf{y},\mathcal{A},\alpha
,\lambda,\eta)$, we consider estimating the label $\tilde{y}$ of a new
observation $\tilde{x}$ using either a point estimate (a single
Bayesian decision list) or the posterior predictive distribution. Given
a point estimate of the antecedent list $d$, we have that
\begin{eqnarray*}
p(\tilde{y}=l | \tilde{x},d,\mathbf{x},\mathbf{y},\alpha) & = &\int
_{\theta} \theta_l p(\theta| \tilde{x},d,\mathbf{x},
\mathbf{y},\alpha) \,d\theta
\\
& = &\mathbb{E}[\theta_l | \tilde{x},d,\mathbf{x},\mathbf{y},
\alpha].
\end{eqnarray*}
Let $j(d,\tilde{x})$ be the index of the first antecedent in $d$ that
applies to $\tilde{x}$. The posterior consequent distribution is
%
\begin{equation}
\label{eq:posterior_consequent} \theta| \tilde{x},d,\mathbf{x},\mathbf{y},\alpha \sim\operatorname
{Dirichlet} (\bolds{\alpha}+\mathbf{N}_{j(d,\tilde{x})} ).
\end{equation}
Thus,
\[
p(\tilde{y}=l | \tilde{x},d,\mathbf{x},\mathbf{y},\alpha) = \frac{\alpha
_l + N_{j(d,\tilde{x}),l}}{\sum_{k=1}^L  ( \alpha_k + N_{j(d,\tilde
{x}),k}  )}.
\]
Additionally, (\ref{eq:posterior_consequent}) allows for the estimation
of 95\% credible intervals using the Dirichlet distribution function.

The posterior mean is often a good choice for a point estimate, but the
interpretation of ``mean'' here is not clear since the posterior is a
distribution over antecedent lists. We thus look for an antecedent list
whose statistics are similar to the posterior mean statistics.
Specifically, we are interested in finding a point estimate $\hat{d}$
whose length $m$ and whose average antecedent cardinality $\bar{c} =
\frac{1}{m} \sum_{j=1}^m c_j$ are close to the posterior mean list
length and average cardinality. Let $\bar{m}$ be the posterior mean
decision list length and\hspace*{1pt} $\bar{\hspace*{-1pt}\bar{c}}$ the posterior mean average
antecedent cardinality, as estimated from the MCMC samples. Then, we
choose our point estimate $\hat{d}$ as the list with the highest
posterior probability among all samples with $m \in\{ \lfloor{\bar
{m}} \rfloor,  \lceil{\bar{m}} \rceil\}$ and
$\bar{c} \in[ \lfloor{\bar{\hspace*{-1pt}\bar{c}}} \rfloor,  \lceil
{\bar{\hspace*{-1pt}\bar{c}}} \rceil]$. We call this point estimate \textit{BRL-point}.

Another possible point estimate is the decision list with the highest
posterior probability---the maximum {a posteriori} estimate.
Given two list lengths, there are many more possible lists of the
longer length than of the shorter length, so prior probabilities in
(\ref{eq:priorprob}) are generally higher for shorter lists. The
maximum {a posteriori} estimate might yield a list that is much
shorter than the posterior mean decision list length, so we prefer the
BRL-point.

In addition to point estimates, we can use the entire posterior
$p(d|\mathbf{x},\mathbf{y},\mathcal{A},\break \alpha,\lambda, \eta)$ to
estimate $y$. The posterior predictive distribution for $y$ is
\begin{eqnarray*}
p(y=l|x,\mathbf{x},\mathbf{y},\mathcal{A},\alpha,\lambda,\eta) &=& \sum
_{d \in\mathbf{D}} p(y=l|x,d,\mathbf{x},\mathbf{y},\mathcal{A},\alpha) p(d|
\mathbf{x},\mathbf{y},\mathcal{A},\alpha,\lambda,\eta)
\\
&= &\sum_{d \in\mathbf{D}} \frac{\alpha_l + N_{j(d,x),l}}{\sum_{k=1}^L
 ( \alpha_k + N_{j(d,x),k}  )} p(d|\mathbf{x},
\mathbf {y},\mathcal{A},\alpha,\lambda,\eta),
\end{eqnarray*}
where $\mathbf{D}$ is the set of all ordered subsets of $\mathcal{A}$.
The posterior samples obtained by MCMC simulation, after burn-in, can
be used to approximate this sum. We call the classifier that uses the
full collection of posterior samples \textit{BRL-post}. Using the
entire posterior distribution to make a prediction means the classifier
is no longer interpretable. One could, however, use the posterior
predictive distribution to classify, and then provide several point
estimates from the posterior to the user as example explanations for
the prediction.

\section{Simulation studies}

We use simulation studies and a deterministic data set to show that
when data are generated by a decision list model, the BRL (Bayesian
Rule Lists; see Section~\ref{sec:intro}) method is able to recover the
true decision list.

\subsection{Simulated data sets}
Given observations with arbitrary features and a collection of rules on
those features, we can construct a binary matrix where the rows
represent observations and the columns represent rules, and the entry
is $1$ if the rule applies to that observation and $0$ otherwise. We
need only simulate this binary matrix to represent the observations
without losing generality. For our simulations, we generated
independent binary rule sets with $100$ rules by setting each feature
value to $1$ independently with probability $1/2$.

We generated a random decision list of size $5$ by selecting 5 rules at
random, and adding the default rule. Each rule in the decision list was
assigned a consequent distribution over labels using a random draw from
the $\operatorname{Beta}(1/2,1/2)$ distribution, which ensures that the rules
are informative about labels. Labels were then assigned to each
observation using the decision list: For each observation, the label
was taken as a draw from the label distribution corresponding to the
first rule that applied to that observation.


For each number of observations $N \in\{100, 250, 500, 1000, 2500,
5000\}$, we generated $100$ independent data sets $(\mathbf{x},\mathbf
{y})$, for a total of $600$ simulated data sets. We did MCMC sampling
with three chains as described in Section~\ref{sec:method} for each
data set. For all data sets, 20,000 samples were sufficient for the
chains to converge.

\begin{figure}

\includegraphics{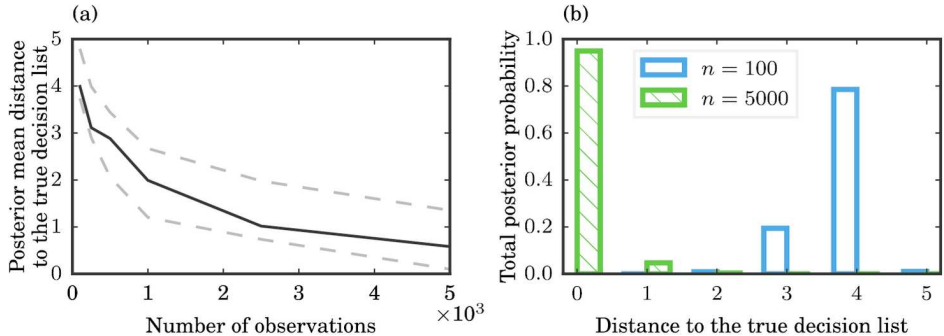}

\caption{\textup{(a)} Average Levenshtein distance from posterior samples to the
true decision list, for differing numbers of observations. The black
solid line indicates the median value across the $100$ simulated data
sets of each size, and the gray dashed lines indicate the first and
third quartiles. \textup{(b)} The proportion of posterior samples with the
specified distance to the true decision list, for a randomly selected
simulation with $n=100$ observations and a randomly selected simulation
with $n=5000$.}
\label{fig:simulations}
\end{figure}

To appropriately visualize the posterior distribution, we binned the
posterior antecedent lists according to their distance from the true
antecedent list, using the Levenshtein string edit distance [\citet
{levenshtein1966}] to measure the distance between two antecedent
lists. This metric measures the minimum number of antecedent
substitutions, additions or removals to transform one decision list
into the other. The results of the simulations are given in Figure~\ref{fig:simulations}.

Figure~\ref{fig:simulations}(a) shows that as the number of
observations increases, the posterior mass concentrates on the true
decision list. Figure~\ref{fig:simulations}(b) illustrates this
concentration with two choices of the distribution of posterior
distances to the true decision list, for $n$ small and for $n$ large.

\subsection{A deterministic problem}\label{sec:tictactoe}
We fit BRL to the Tic--Tac--Toe Endgame data set from the UCI Machine
Learning Repository [\citet{BacheLichman:2013}] of benchmark data sets.
The Tic--Tac--Toe Endgame data set provides all possible end board
configurations for the game Tic--Tac--Toe, with the task of determining
if player ``X'' won or not. The data set is deterministic, with exactly
8 ways that player ``X'' can win, each one of the 8 ways to get 3 ``X'''s
in a row on a $3\times 3$ grid. We split the data set into 5 folds and did
cross-validation to estimate test accuracy. For each fold of
cross-validation, we fit BRL with prior hyperparameters $\lambda= 8$
and $\eta= 3$, and the point estimate decision list contained the 8
ways to win and thus achieved perfect accuracy. In Table~\ref{table:uci_accuracy}, we compare accuracy on the test set with C5.0,
CART, $\ell_1$-regularized logistic regression ($\ell_1$-LR), RBF
kernel support vector machines (SVM), random forests (RF) and Bayesian
CART (BCART). The implementation details for these comparison
algorithms are in the \hyperref[app]{Appendix}. None of these other methods was able to
achieve perfect accuracy. Decision trees in particular are capable of
providing a perfect classifier for this problem, but the greedy
learning done by C5.0 and CART did not find the perfect classifier.

\begin{table}
\caption{Mean classification accuracy in the top row, with standard
deviation in the second row, for machine learning algorithms using 5
folds of cross-validation on the Tic--Tac--Toe Endgame data set}\label{table:uci_accuracy}
\begin{tabular*}{\textwidth}{@{\extracolsep{\fill}}lccccccc@{}}
\hline
& \textbf{BRL} & \textbf{C5.0} & \textbf{CART} & $\bolds{\ell_1}$\textbf{-LR} & \textbf{SVM} & \textbf{RF} & \textbf{BCART} \\
\hline
Mean accuracy & 1.00 & 0.94 & 0.90 & 0.98 & 0.99 & 0.99 & 0.71 \\
Standard deviation & 0.00 & 0.01 & 0.04 & 0.01 & 0.01 & 0.01 & 0.04\\
\hline
\end{tabular*}
\end{table}

\section{Stroke prediction}\label{sec:chads2}
We used Bayesian Rule Lists to derive a stroke prediction model using
the MarketScan Medicaid Multi-State Database (MDCD). MDCD contains
administrative claims data for 11.1 million Medicaid enrollees from
multiple states. This database forms part of the suite of databases
from the Innovation in Medical Evidence Development and Surveillance
(IMEDS, \url{http://imeds.reaganudall.org/}) program that have been\break 
mapped to a common data model [\citet{Stang10}].

We extracted every patient in the MDCD database with a diagnosis of
atrial fibrillation, one year of observation time prior to the
diagnosis and one year of observation time following the diagnosis
($n=12\mbox{,}586$). Of these, 1786 (14\%) had a stroke within a year of the
atrial fibrillation diagnosis.

As candidate predictors, we considered all drugs and all conditions.
Specifically, for every drug and condition, we created a binary
predictor variable indicating the presence or absence of the drug or
condition in the full longitudinal record prior to the atrial
fibrillation diagnosis. These totaled 4146 unique medications and
conditions. We included features for age and gender. Specifically, we
used the natural values of 50, 60, 70 and 80 years of age as split
points, and for each split point introduced a pair of binary variables
indicating if age was less than or greater than the split point.
Considering both patients and features, here we apply our method to a
data set that is over 6000 times larger than that originally used to
develop the CHADS$_2$ score (which had $n=1733$ and considered 5 features).

We did five folds of cross-validation. For each fold, we pre-mined the
collection of possible antecedents using frequent itemset mining with a
minimum support threshold of $10\%$ and a maximum cardinality of $2$.
The total number of antecedents used ranged from $2162$ to $2240$
across the folds. We set the antecedent list prior hyperparameters
$\lambda$ and $\eta$ to $3$ and $1$, respectively, to obtain a Bayesian
decision list of similar complexity to the CHADS$_2$ score. For each
fold, we evaluated the performance of the BRL point estimate by
constructing a receiver operating characteristic (ROC) curve and
measuring area under the curve (AUC) for each fold.

In Figure~\ref{fig:mdcd_declist} we show the BRL point estimate
recovered from one of the folds. The list indicates that past history
of stroke reveals a lot about the vulnerability toward future stroke.
In particular, the first half of the decision list focuses on a history
of stroke, in order of severity. Hemiplegia, the paralysis of an entire
side of the body, is often a result of a severe stroke or brain injury.
Cerebrovascular disorder indicates a prior stroke, and transient
ischaemic attacks are generally referred to as ``mini-strokes.'' The
second half of the decision list includes age factors and vascular
disease, which are known risk factors and are included in the
CHA$_2$DS$_2$-VASc score. The BRL-point lists that we obtained in the 5
folds of cross-validation were all of length 7, a similar complexity to
the CHADS$_2$ and CHA$_2$DS$_2$-VASc scores which use 5 and 8 features,
respectively.

The point estimate lists for all five of the folds of cross-validation
are given in the supplemental material [\citet{supp}]. There is significant overlap in
the antecedents in the point estimates across the folds. This suggests
that the model may be more stable in practice than decision trees,
which are notorious for producing entirely different models after small
changes to the training set [\citeauthor{Breiman962}
(\citeyear{Breiman962,Breiman96})].

\begin{figure}
\begin{framed}
\begin{flushleft}
\textbf{if} hemiplegia \textbf{and} age${}>{}$60 \textbf{then}
\textit{stroke risk} 58.9\% (53.8\%--63.8\%)\\
\textbf{else if} cerebrovascular disorder \textbf{then} \textit{stroke
risk} 47.8\% (44.8\%--50.7\%)\\
\textbf{else if} transient ischaemic attack \textbf{then} \textit
{stroke risk} 23.8\% (19.5\%--28.4\%)\\
\textbf{else if} occlusion and stenosis of carotid artery without
infarction \textbf{then} \textit{stroke risk} 15.8\% (12.2\%--19.6\%)\\
\textbf{else if} altered state of consciousness \textbf{and} age${}>{}$60
\textbf{then} \textit{stroke risk} 16.0\% (12.2\%--20.2\%)\\
\textbf{else if} age${}\leq{}$70 \textbf{then} \textit{stroke risk} 4.6\%
(3.9\%--5.4\%)\\
\textbf{else} \textit{stroke risk} 8.7\% (7.9\%--9.6\%)
\end{flushleft}
\end{framed}
\caption{Decision list for determining 1-year stroke risk following
diagnosis of atrial fibrillation from patient medical history. The risk
given is the mean of the posterior consequent distribution, and in
parentheses is the 95\% credible interval.}
\label{fig:mdcd_declist}
\end{figure}

\begin{figure}

\includegraphics{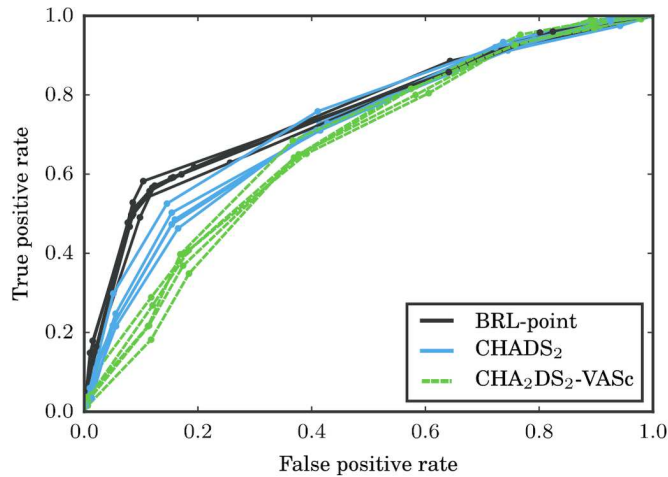}

\caption{ROC curves for stroke prediction on the MDCD database for each
of 5 folds of cross-validation, for the BRL point estimate, CHADS$_2$
and CHA$_2$DS$_2$-VASc.}
\label{fig:mdcd_roc}
\end{figure}

\begin{table}[b]
\caption{Mean, and in parentheses standard deviation, of AUC and
training time across 5 folds of cross-validation for stroke prediction.
Note that the CHADS$_2$ and CHA$_2$DS$_2$-VASc models are fixed, so no
training time is reported}\label{table:mdcd_auc}
\begin{tabular*}{\textwidth}{@{\extracolsep{\fill}}lcd{3.9}@{}}
\hline
& \textbf{AUC} & \multicolumn{1}{c@{}}{\textbf{Training time (mins)}} \\
\hline
BRL-point& 0.756 (0.007) & 21.48\ (6.78) \\
CHADS$_2$& 0.721 (0.014) & \multicolumn{1}{c}{no training} \\
CHA$_2$DS$_2$-VASc& 0.677 (0.007) & \multicolumn{1}{c}{no training} \\
CART & 0.704 (0.010) & 12.62\ (0.09) \\
C5.0 & 0.704 (0.011) & 2.56\ (0.27) \\
$\ell_1$ logistic regression & 0.767 (0.010) & 0.05\ (0.00) \\
SVM& 0.753 (0.014) & 302.89\ (8.28) \\
Random forests& 0.774 (0.013) & 698.56\ (59.66)\\
BRL-post& 0.775 (0.015) & 21.48\ (6.78)\\
\hline
\end{tabular*}
\end{table}

In Figure~\ref{fig:mdcd_roc} we give ROC curves for all 5 folds for
BRL-point, CHADS$_2$ and CHA$_2$DS$_2$-VASc, and in Table~\ref{table:mdcd_auc} we report mean AUC across the folds. These results
show that with complexity and interpretability similar to CHADS$_2$,
the BRL point estimate decision lists performed significantly better at
stroke prediction than both CHADS$_2$ and CHA$_2$DS$_2$-VASc.
Interestingly, we also found that CHADS$_2$ outperformed
CHA$_2$DS$_2$-VASc despite CHA$_2$DS$_2$-VASc being an extension of
CHADS$_2$. This is likely because the model for the CHA$_2$DS$_2$-VASc
score, in which risk factors are added linearly, is a poor model of
actual stroke risk. For instance, the stroke risks estimated by
CHA$_2$DS$_2$-VASc are not a monotonic function of score. Within the
original CHA$_2$DS$_2$-VASc calibration study, \citet{Lip10} estimate a
stroke risk of 9.6\% with a CHA$_2$DS$_2$-VASc score of 7, and a 6.7\%
risk with a score of 8. The indication that more stroke risk factors
can correspond to a lower stroke risk suggests that the
CHA$_2$DS$_2$-VASc model may be misspecified, and highlights the
difficulty in constructing these interpretable models manually.

The results in Table~\ref{table:mdcd_auc} give the AUC for BRL,
CHADS$_2$, CHA$_2$DS$_2$-VASc, along with the same collection of
machine learning algorithms used in Section~\ref{sec:tictactoe}. The
decision tree algorithms CART and C5.0, the only other interpretable
classifiers, were outperformed even by CHADS$_2$. The BRL-point
performance was comparable to that of SVM, and not substantially worse
than $\ell_1$ logistic regression and random forests. Using the full
posterior, BRL-post matched random forests for the best performing method.

All of the methods were applied to the data on the same, single Amazon
Web Services virtual core with a processor speed of approximately
2.5~GHz and 4~GB of memory. Bayesian CART was unable to fit the data
since it ran out of memory, and so it is not included in Table~\ref{table:mdcd_auc}.

The BRL MCMC chains were simulated until convergence, which required
50,000 iterations for 4 of the 5 folds, and 100,000 for the fifth. The
three chains for each fold were simulated in serial, and the total CPU
time required per fold is given in Table~\ref{table:mdcd_auc}, together
with the CPU times required for training the comparison algorithms on
the same processor. Table~\ref{table:mdcd_auc} shows that the BRL MCMC
simulation was more than ten times faster than training SVM, and more
than thirty times faster than training random forests, using standard
implementations of these methods as described in the \hyperref[app]{Appendix}.

\subsection{Additional experiments}
We further investigated the properties and performance of the BRL by
applying it to two subsets of the data, female patients only and male
patients only. The female data set contained 8368 observations, and the
number of pre-mined antecedents in each of 5 folds ranged from 1982 to
2197. The male data set contained 4218 observations, and the number of
pre-mined antecedents in each of 5 folds ranged from 1629 to 1709. BRL
MCMC simulations and comparison algorithm training were done on the
same processor as the full experiment. The AUC and training time across
five folds for each of the data sets is given in Table~\ref{table:mdcd_otherexp_auc}.

The BRL point estimate again outperformed the other interpretable
models (CHADS$_2$, CHA$_2$DS$_2$-VASc, CART and C5.0), and the BRL-post
performance matched that of random forests for the best performing
method. As before, BRL MCMC simulation required significantly less time
than SVM or random forests training. Point estimate lists for these
additional experiments are given in the supplemental materials [\citet{supp}].

\begin{table}
\caption{Mean, and in parentheses standard deviation, of AUC and
training time (mins) across 5 folds of cross-validation for stroke prediction}\label{table:mdcd_otherexp_auc}
\begin{tabular*}{\textwidth}{@{\extracolsep{\fill}}lcd{3.9}cd{3.9}@{}}
\hline
& \multicolumn{2}{c}{\textbf{Female patients}} &
\multicolumn{2}{c@{}}{\textbf{Male patients}} \\[-6pt]
& \multicolumn{2}{c}{\hrulefill} &
\multicolumn{2}{c@{}}{\hrulefill} \\
& \textbf{AUC} & \multicolumn{1}{c}{\textbf{Training time}} & \textbf{AUC} & \multicolumn{1}{c@{}}{\textbf{Training time}}\\
\hline
BRL-point& 0.747 (0.028) & 9.12\ (4.70) & 0.738 (0.027)& 6.25\ (3.70)\\
CHADS$_2$& 0.717 (0.018) & \multicolumn{1}{c}{no training} & 0.730 (0.035) & \multicolumn{1}{c}{no training}\\
CHA$_2$DS$_2$-VASc& 0.671 (0.021) & \multicolumn{1}{c}{no training} & 0.701 (0.030) & \multicolumn{1}{c}{no
training}\\
CART & 0.704 (0.024) & 7.41\ (0.14) & 0.581 (0.111) & 2.69\ (0.04)\\
C5.0 & 0.707 (0.023) & 1.30\ (0.09) & 0.539 (0.086) & 0.55\ (0.01)\\
$\ell_1$ logistic regression & 0.755 (0.025) & 0.04\ (0.00) & 0.739
(0.036)& 0.01\ (0.00)\\
SVM& 0.739 (0.021) & 56.00\ (0.73) & 0.753 (0.035) & 11.05\ (0.18) \\
Random forests& 0.764 (0.022) & 389.28\ (33.07) & 0.773 (0.029) & 116.98\
(12.12) \\
BRL-post& 0.765 (0.025) & 9.12\ (4.70) & 0.778 (0.018) & 6.25\ (3.70)\\
\hline
\end{tabular*}
\end{table}

\section{Related work and discussion}
Most widely used medical scoring systems are designed to be
interpretable, but are not necessarily optimized for accuracy, and
generally are derived from few factors. The Thrombolysis In Myocardial
Infarction (TIMI) Score [\citet{AntmanEtAl}], Apache II score for infant
mortality in the ICU [\citet{apacheII}], the CURB-65 score for
predicting mortality in community-acquired pneumonia [\citet
{Lim01052003}] and the CHADS$_2$ score [\citet{Gage:2001ed}] are
examples of interpretable predictive models that are very widely used.
Each of these scoring systems involves very few calculations and could
be computed by hand during a doctor's visit. In the construction of
each of these models, heuristics were used to design the features and
coefficients for the model; none of these models was fully learned from data.

In contrast with these hand-designed interpretable medical scoring
systems, recent advances in the collection and storing of medical data
present unprecedented opportunities to develop powerful models that can
predict a wide variety of outcomes [\citet{Schmueli10}].
The front-end user interface of medical risk assessment tools are
increasingly available online (e.g., \url{http://www.r-calc.com}). At
the end of the assessment, a patient may be told he or she has a high
risk for a particular outcome but without understanding why the
predicted risk is high, particularly if many pieces of information were
used to make the prediction.

In general, humans can handle only a handful of cognitive entities at
once [\citet{Miller56,Jennings82}]. It has long since been hypothesized
that simple models predict well, both in the machine learning
literature [\citet{Holte93}] and in the psychology literature [\citet
{dawes1979robust}]. The related concepts of explanation and
comprehensibility in statistical modeling have been explored in many
past works [\citet
{bratko1997machine,Madigan97,Giraud98,ruping2006learning,Huysmans11,VellidoEtAl12,Freitas14},
e.g.].

Decision lists have the same form as models used in the expert systems
literature from the 1970s and 1980s [\citet{Leondes02}], which were
among the first successful types of artificial intelligence. The
knowledge base of an expert system is composed of natural language
statements that are \emph{if\ldots then\ldots} rules. Decision lists are a
type of associative classifier, meaning that the list is formed from
association rules. In the past, associative classifiers have been
constructed from heuristic greedy sorting mechanisms [\citet{Rivest87,Liu98,Marchand05,RudinLeMa13}]. Some of these sorting mechanisms work
provably well in special cases, for instance, when the decision problem
is easy and the classes are easy to separate, but are not optimized to
handle more general problems. Sometimes associative classifiers are
formed by averaging several rules together, or having the rules each
vote on the label and then combining the votes, but the resulting
classifier is not generally interpretable [\citet{Li01,Yin03,Friedman08,Meinshausen10}]. 

In a previous paper we proved that the VC dimension of decision list
classifiers equals $|\mathcal{A}|$, the number of antecedents used to
learn the model [Theorem~3, \citet{RudinLeMa13}]. This result leads to a
uniform generalization bound for decision lists [Corollary~4, \citet
{RudinLeMa13}]. This is the same as the VC dimension obtained by using
the antecedents as features in a linear model, thus we have the same
prediction guarantees. We then expect similar generalization behavior
for decision lists and weighted linear combination models.

BRL interacts with the feature space only through the collection of
antecedents $\mathcal{A}$. The computational effort scales with the
number of antecedents, not the number of features, meaning there will
generally be less computation when the data are sparse. This means that
BRL tends to scale with the sparsity of the data rather than the number
of features.

Decision trees are closely related to decision lists, and are in some
sense equivalent: any decision tree can be expressed as a decision
list, and any decision list is a one-sided decision tree. Decision
trees are almost always constructed greedily from the top down, and
then pruned heuristically upward and cross-validated to ensure accuracy.
Because the trees are not fully optimized, if the top of the decision
tree happened to have been chosen badly at the start of the procedure,
it could cause problems with both accuracy and interpretability.
Bayesian decision trees [\citeauthor{Chipman:1998jh}
(\citeyear{Chipman:1998jh,Chipman:2002hc}),
\citet{Dension:1998hl}] use Markov chain Monte Carlo (MCMC) to sample from a
posterior distribution over trees. Since they were first proposed,
several improvements and extensions have been made in both sampling
methods and model structure [\citet{Wu:2007dh,Chipman10,Taddy11}]. The
space of decision lists using pre-mined rules is significantly smaller
than the space of decision trees, making it substantially easier to
obtain MCMC convergence and to avoid the pitfalls of local optima.
Moreover, rule mining allows for the rules to be individually powerful.
Constructing a single decision tree is extremely fast, but sampling
over the space of decision trees is extremely difficult (unless one is
satisfied with local maxima). To contrast this with our approach, the
rule mining step is extremely fast, yet sampling over the space of
decision lists is very practical.

There is a subfield of artificial intelligence, Inductive Logic
Programming [\citet{muggleton1994inductive}], whose goal is to mine
individual conjunctive rules. It is possible to replace the frequent
itemset miner with an inductive logic programming technique, but this
generally leads to losses in predictive accuracy; ideally, we would use
a large number of diverse rules as antecedents, rather than a few
(highly overlapping) complex rules as would be produced by an ILP
algorithm. In our experiments to a follow-up work [\citet{WangRu15}],
the use of an ILP algorithm resulted in a substantial loss in performance.

Interpretable models are generally not unique (stable), in the sense
that there may be many equally good models, and it is not clear in
advance which one will be returned by the algorithm. For most problems,
the space of high quality predictive models is fairly large [called the ``Rashomon Effect'' \citet{breiman2001}],
so we cannot expect
uniqueness. In practice, as we showed, the rule lists across test folds
were very similar, but if one desires stability to small perturbations
in the data generally, we recommend using the full posterior rather
than a point estimate. The fact that many high performing rule lists
exist can be helpful, since it means the user has many choices of which
model to use.

This work is related to the Hierarchical Association Rule Model (HARM),
a~Bayesian model that uses rules [\citet{McCormick:2011ws}]. HARM
estimates the conditional probabilities of each rule jointly in a
conservative way. Each rule acts as a separate predictive model, so
HARM does not explicitly aim to learn an ordering of rules.

There are related works on learning decision lists from an optimization
perspective. In particular, the work of \citet{RudinEr15} uses
mixed-integer programming to build a rule list out of association
rules, which has guarantees on optimality of the solution. Similarly to
that work, \citet{GohRu14} fully learn sparse disjunctions of
conjunctions using optimization methods.

There have been several follow-up works that directly extend and apply
Bayesian Rule Lists.
The work of \citet{WangRu15} on Falling Rule Lists provides a
nontrivial extension to BRL whereby the probabilities for the rules are
monotonically decreasing down the list. \citet{WangEtAl15} build
disjunctions of conjunctive rules using a Bayesian framework similar to
the one in this work.
\citet{ZhangEtAl15} have taken an interesting approach to constructing
optimal treatment regimes using a BRL-like method, where, in addition
to the criteria of accuracy, the rule list has a decision cost for
evaluating it. It is possible to use BRL itself for that purpose as
well, as one could give preference to particular antecedents that cost
less. This sort of preference could be expressed in the antecedent
prior distribution in (\ref{eq:uniformprior}).
\citet{King14} have taken a Bayesian Rule List approach to handle a
challenging problem in text analysis, which is to build a keyword-based
classifier that is easier to understand in order to solicit high
quality human input. \citet{Souillard15} applied Bayesian Rule Lists
and Falling Rule Lists to the problem of screening for cognitive
disorders such as Alzheimer's disease based on the digitized pen
strokes of patients during the Clock Drawing test.

Shorter preliminary versions of this work are those of
\citeauthor{LethamRuMcMaAAAI13}
(\citeyear
{LethamRuMcMaAAAI13,LethamRuMcMa14}). \citet
{LethamRuMcMaAAAI13} used a different prior and called the algorithm
the Bayesian List Machine.

\section{Conclusion}\label{sec:disc}
We are working under the hypothesis that many real data sets permit
predictive models that can be surprisingly small. This was hypothesized
over two decades decade ago [\citet{Holte93}]; however, we now are starting to
have the computational tools to truly test this hypothesis. The BRL
method introduced in this work aims to hit the ``sweet spot'' between
predictive accuracy, interpretability and tractability.

Interpretable models have the benefits of being both concise and
convincing. A~small set of trustworthy rules can be the key to
communicating with domain experts and to allowing machine learning
algorithms to be more widely implemented and trusted. In practice, a
preliminary interpretable model can help domain experts to troubleshoot
the inner workings of a complex model, in order to make it more
accurate and tailored to the domain. We demonstrated that interpretable
models lend themselves to the domain of predictive medicine, and there
is a much wider variety of domains in science, engineering and
industry, where these models would be a natural choice.

\begin{appendix}\label{app}

\section*{Appendix}
\subsection*{Comparison algorithm implementations}
\textit{Support vector machines}: LIBSVM [\citet{Chang01}] with
a radial basis function kernel. We selected the slack parameter
$C_{\mathrm{SVM}}$ and the kernel parameter $\gamma$ using a grid
search over the ranges $C_{\mathrm{SVM}} \in\{2^{-2},2^{0},\ldots
,2{^6}\}$ and $\gamma\in\{2^{-6}, 2^{-4}, \ldots, 2^2\}$. We chose
the set of parameters with the best 3-fold cross-validation performance
using LIBSVM's built-in cross-validation routine. \textit{C5.0}: The R
library ``C50'' with default settings. \textit{CART}: The R library
``rpart'' with default parameters and pruned using the complexity
parameter that minimized cross-validation error. \textit{Logistic
regression}: The LIBLINEAR [\citet{Fan08}] implementation of logistic
regression with $\ell_1$ regularization. We selected the regularization
parameter $C_{\mathrm{LR}}$ from $\{2^{-6},2^{-4},\ldots,2^6\}$ as that
with the best 3-fold cross-validation performance, using LIBLINEAR's
built-in cross-validation routine. \textit{Random forests}: The R
library ``randomForest.'' The optimal value for the parameter ``mtry''
was found using ``tuneRF,'' with its default 50 trees. The optimal
``mtry'' was then used to fit a random forests model with 500 trees, the
library default. \textit{Bayesian CART}: The R library ``tgp,'' function
``bcart'' with default settings.
\end{appendix}

\section*{Acknowledgments}
The authors thank Zachary Shahn and the OMOP team for help with the data.

\begin{supplement}
\stitle{Computer code}
\slink[doi]{10.1214/15-AOAS848SUPPA} 
\sdatatype{.zip}
\sfilename{aoas848\_suppa.zip}
\sdescription{Our Python code used to fit decision lists to data, along
with an example data set.}
\end{supplement}

\begin{supplement}
\stitle{BRL point estimates}
\slink[doi]{10.1214/15-AOAS848SUPPB} 
\sdatatype{.pdf}
\sfilename{aoas848\_suppb.pdf}
\sdescription{The BRL point estimates for all of the cross-validation
folds for the stroke prediction experiment, and BRL-point estimates for
the female-only and male-only experiments.}
\end{supplement}





\printaddresses
\end{document}